\documentstyle[12pt,aaspp]{article}

\newcommand{\km}{km s$^{-1}$}
\newcommand{\vi}{v_{\scriptscriptstyle 0}}    
\newcommand{\vy}[2]{#1_{\scriptscriptstyle #2}}

\def\gtorder{\mathrel{\raise.3ex\hbox{$>$}\mkern-14mu
             \lower0.6ex\hbox{$\sim$}}}
\def\ltorder{\mathrel{\raise.3ex\hbox{$<$}\mkern-14mu
             \lower0.6ex\hbox{$\sim$}}}
\def\proptwid{\mathrel{\raise.3ex\hbox{$\propto$}\mkern-14mu
             \lower0.6ex\hbox{$\sim$}}}
\begin{document}
\title{Are AGN Broad Emission Lines Formed by Discrete Clouds? --- 
Analysis of Keck High Resolution Spectroscopy of NGC 4151}

  
\author{Nahum Arav\footnote{Theoretical Astrophysics, Caltech 130-33,
Pasadena, California 91125, USA \\ 
current address: IGPP, LLNL, L-413, P.O. Box 808, Livermore, CA 94550; 
I: narav@igpp.llnl.gov}, 
 Tomas A.\ Barlow\footnote{Department of Astronomy, Caltech
105-24, Pasadena, California 91125, USA}, 
Ari Laor\footnote{Physics Department, Technion, Haifa 3200, Israel},
   Wallace. L. W. Sargent$^2$ \& Roger D.\ Blandford$^1$ }


\begin{abstract}

We search for a direct signature of discrete ``clouds'' in the
broad line region (BLR) of the Seyfert galaxy NGC 4151. For this purpose 
we apply cross correlation (CC) analysis to a high resolution
KECK spectrum of the galaxy. No such signature is found in the data.
In order for cloud models to be compatible with this result, there
must be at least  $\sim3\times10^7$
emitting clouds in the BLR, 
 where the limit is based on simulation of a homogeneous
cloud population. More realistic distributions increase the
lower limit to above $10^8$.
These numbers are an order of magnitude
improvement on our previous limit from Mrk 335, where the improvement
 comes from
higher S/N, broader lines, and refined simulations. 
Combined with 
the predicted upper limit for the number of emitting clouds in NGC
4151 ($10^6-10^7$), the derived lower limit puts a strong constraint on 
the cloud scenario in the BLR of this object.  
Similar constraints can be paled on models where the emission
originates in streams and sheets.
Thus, this investigation
suggests that the BELs in NGC 4151, and by extension in all
AGNs, are not made of an ensemble of discrete independent
emitters.


\end{abstract}

\section{INTRODUCTION}
Broad emission lines (BELs) are among the most prominent feature
 observed in Active Galactic Nuclei (AGN).
The standard picture is that the BEL material is in the form of dense
clouds with a volume filling factor of $\sim10^{-6}$ illuminated by
the a central ionizing continuum source (Netzer 1990:
Arav et al. 1997).
Two main cloud models appear in the literature: 
 One is the two-phase model in which cool clouds 
($T\sim10^4$ K) are embedded in a hot medium with $T\sim10^8$ K which
confines the clouds
(Krolik, McKee \& Tarter
1981). Another class of models create the BELs out of
stellar atmospheres or ``bloated'' stars (Scoville \& Norman 1988; Kazanas
1989; Alexander \& Netzer 1994). In these models the individual sources
are slow ($\sim10$ \km) outflows emanating from super giant stars. 
For further description of the two-phase and bloated star 
models including some of their shortcomings, see 
Arav et al. (1997) and references therein.
Alternatives to the cloud picture can be found in magnetic driven
winds (Emmering, Blandford, \& Shlosman 1992), and in models of disk
emission coupled with  winds (Chiang \&
Murray 1996).

In a previous paper (Arav et al. 1997) we described an investigation aimed at
testing the cloud picture. Here we give a brief description of the
basic idea and our methodology. An ensemble of discrete emitting
units is expected to give a smooth line profile only when the number
of units approaches infinity. For any finite number of randomly
distributed clouds there will be
fluctuations associated with the discreteness of the profile building
blocks. The same micro structure that is
caused by the fluctuations in one line profile should appear in
different observations of the same line, and in different lines which
arise from the same ion. On the other hand, 
photon shot noise, which is the major
cause for fluctuations (in essence the S/N of the observation),
is a random process, and therefore there is hope to obtain a clear
distinction between the two sources of fluctuation. Furthermore,
since fluctuations due to clouds in two different profiles ought to correlate 
along the entire profile, we should
be able to detect them even if they are locally smaller than the random photon
shot noise.   

The most straightforward way to test for similar microstructure in
different profiles is to use cross-correlation (CC) techniques 
(for details see Arav et al. 1997). 
Such an approach was used by Atwood, Baldwin \&
Carswell  (1982)
on a moderate resolution spectra of Mrk 509 and the lack of a strong CC
at zero velocity shift in
their data was interpreted as a lower limit of $5\times 10^4$ for the
number of emitting clouds ($N_c$).  
By using an improved CC analysis on high quality data from Mrk 335, and
extensive Monte Carlo simulations, we
were able to put a lower limit of $N_c\gtorder3\times10^6$ in this
object.
This limit applies to identical clouds
with realistic temperature (T=2$\times10^4$ K) and optical depth 
($\sim10^4$ in the  H${\alpha}$ line). 
Since the lower limit on $N_c$ is at least 30 times larger than the
maximum  number of stars in the bloated star models, these   
 models can be ruled 
out, unless
the line width of the gas associated with an 
 individual star exceeds the unrealistically large
value of 100 \km. 
Simple photoionization arguments yield estimates for 
the largest possible number
of individual clouds without adhering to a specific cloud model.
Therefore, it is of great interest to try to constrain this 
model-independent estimate.  
In  Mrk 335 the  photoionization
estimate 
 for the upper limit of $N_c$  vary
between $10^7-10^8$ with large uncertainties, and thus is compatible
with our lower limit. A significant improvement in the limit on $N_c$
is needed in order to  confront the photoionization
estimate. 

 To achieve this goal we observed NGC 4151 and analyzed its BELs. 
For our purposes this object has three advantages 
 compared with  Mrk 335:
1) Reverberation studies (Maoz et al. 1991; see also a more recent
campaign: Kaspi et al. 1996; Crenshaw et al. 1996; Edelson et al. 1996)
 indicate a BLR size of roughly 9 light days. Substituting this number
to our estimate of $N_c$ (Arav et al. 1997) we obtain an upper limit
 of $10^7$ clouds (or $10^6$ using the estimate found in Netzer 1990) 
in NGC 4151, which is an order of magnitude {\it smaller} than  
 the one in Mrk 335.  These estimates are 
$\propto U^{-4}$, where $U$ is the ionization  
parameter, and the above upper limits are for the commonly used value
of $U=0.1$.
2) In spite of being less luminous, the closer distance of NGC 4151 makes
it roughly two magnitudes brighter than Mrk 335. The resultant higher
S/N increases the Monte Carlo derived lower limit for 
$N_c$.
3) The BELs in NGC 4151 are three times wider than in Mrk 335. With
all else equal, three times more clouds are needed to cover a three times
wider line. 
Advantages 2 and 3, combined with improvements in our algorithms,
increase the lower limit on $N_c$ to $3\times10^7$, an order of
magnitude {\it larger} than the one for Mrk 335. Since the photoionization
estimate for the upper limit on $N_c$ is an order of magnitude lower,
the combined effect is a two orders of magnitude improvement 
in the constraint on this model
independent estimate.  

A possible caveat comes from the fact 
 that we used clouds with a number density of $\simeq 10^{10}$ cm$^{-2}$
 in our upper limit
estimates on $N_c$.  This $\vy{n}{H}$ is assumed since the 
 C~III]~$\lambda$1909 BEL
is strongly quenched at higher densities.  However,  models
where a significant fraction of the Balmer lines emission comes from a
higher density gas were proposed in the literature (Ferland \& Rees 1988).
The upper limit on $N_c$ for these models will be
substantially higher. In order to test 
these models a  similar experiment to the one we show here 
should be performed on the C~III]~$\lambda$1909 BEL in NGC 4151, for
which the upper limit
estimates on $N_c$ should hold for all models.
This will require high resolution spectroscopy with the HST.     
 
In this paper we describe the NGC 4151 observations (\S~2), present
the results of
the CC analysis (\S~3), describe our Monte Carlo simulations from
which we obtain the lower limits on
$N_c$ for different emission entities (\S~4), and discuss the
implications of these result (\S~4).

\section{DATA ACQUISITION AND REDUCTION}

\begin{figure}
\plotone{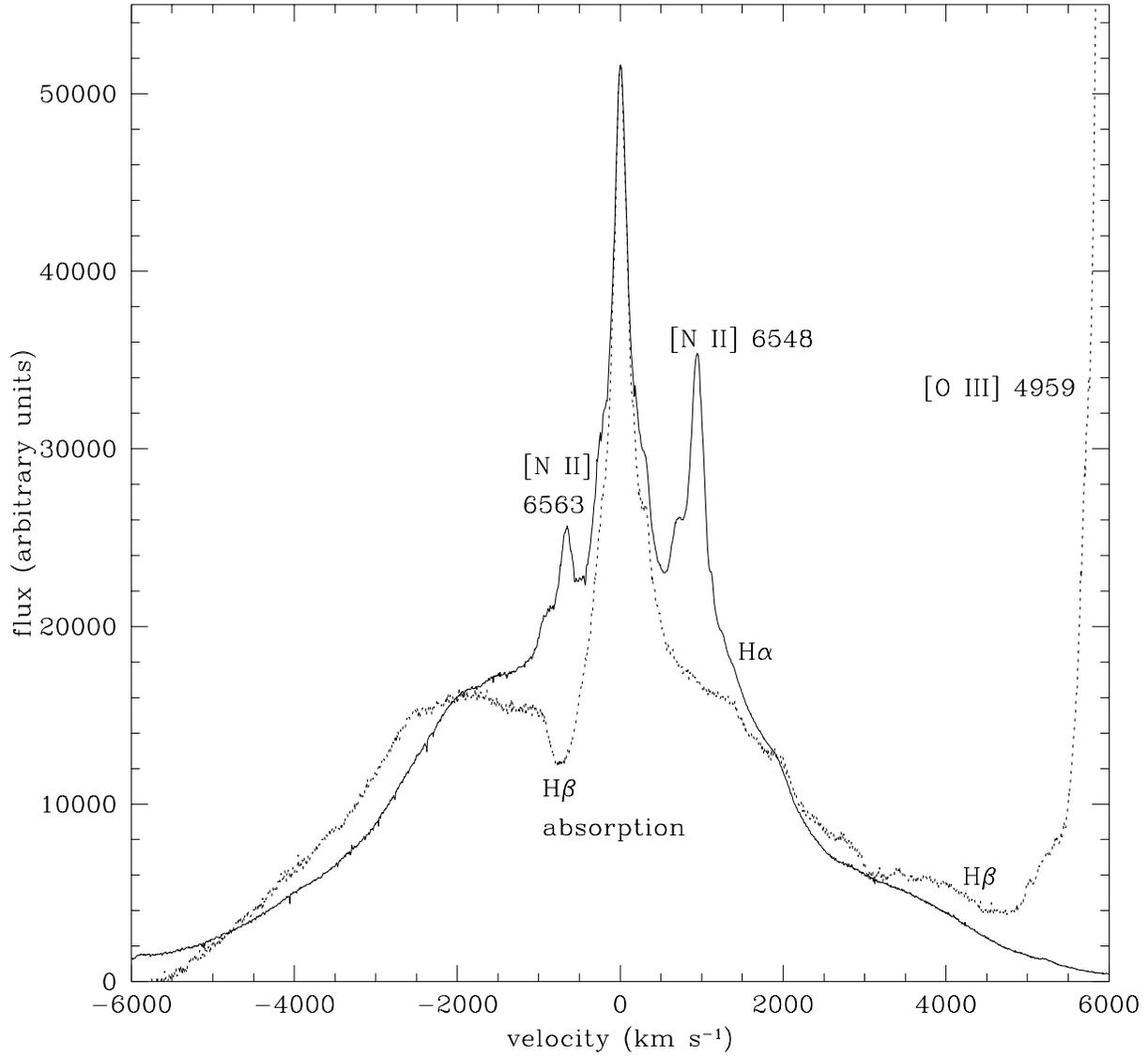}
\caption{Observed profiles of the 
H${\alpha}$ (solid line) and H${\beta}$ (dotted line)
lines in  NGC
4151 (continuum
subtracted), in arbitrary scaling that matches
the peaks of the narrow line components. Some associated feature are
also labeled.} \label{fig-1}
\end{figure}

The HIRES spectra for the Seyfert galaxy NGC 4151 were obtained
with the Keck I telescope on May 22, 1996.  We acquired 4 exposures
totaling 56 minutes. For the purpose of our analysis two exposures
suffice but by acquiring four we reduce the systematic errors in the data.
 Each exposure used a 0.86 arc-second slit (3
pixels) for a resolution of approximately 6.3 \km.  The actual
resolution varies from about 6.0 to 6.5 km/s FWHM due to the change in
dispersion across each 4200 \km\ wide echelle order.
The wavelength coverage was 4400 to 6800 \AA\ with some gaps
beyond 5000 \AA.  The signal-to-noise per (3 pixel) resolution
element in the total exposure 
ranged from about 300 in the continuum up to $\sim$800 and
$\sim$400 for the peak of the BEL component of the H$\alpha$ and
H$\beta$ lines, respectively.
The wavelength calibration error is about 0.1 pixel (0.2 \km)
root-mean-square and up to a maximum error of about 0.3 pixels (0.6
\km).  This is the relative error across the spectrum.  The absolute
wavelength error is about 0.3 pixels (0.6 \km)
 So altogether
there can be up to $\sim$1 km/s error in the absolute wavelength value
at any given pixel.

Severe fluxing errors occur when trying to combine echelle orders into
a single continuous spectrum.  This problem was exasperated by the
narrow-line region emission which has an angular extent larger than the
1 arc-second FWHM of the continuum spatial profile.  As a result, the
shape of structures larger than about 20 angstroms are strongly
affected by the rescaling process necessary to average the exposures.
Relative flux level errors of 5 to 20 percent may occur across sections
of spectrum larger than 20 angstroms.
The data was reduced using our own reduction package (Barlow and
Sargent 1997) which automatically traces the orders and ignores
anomalous radiation events in the sky region.  Since the narrow
emission-line region is resolved, we could not use profile
modeling to correct for cosmic rays in the object.  The cosmic rays
were corrected after extracting the spectra and are evident as
``spikes'' in the error spectrum.
Figure 1 shows the observed H${\alpha}$ and H${\beta}$ line profiles.

\begin{figure}
\plotone{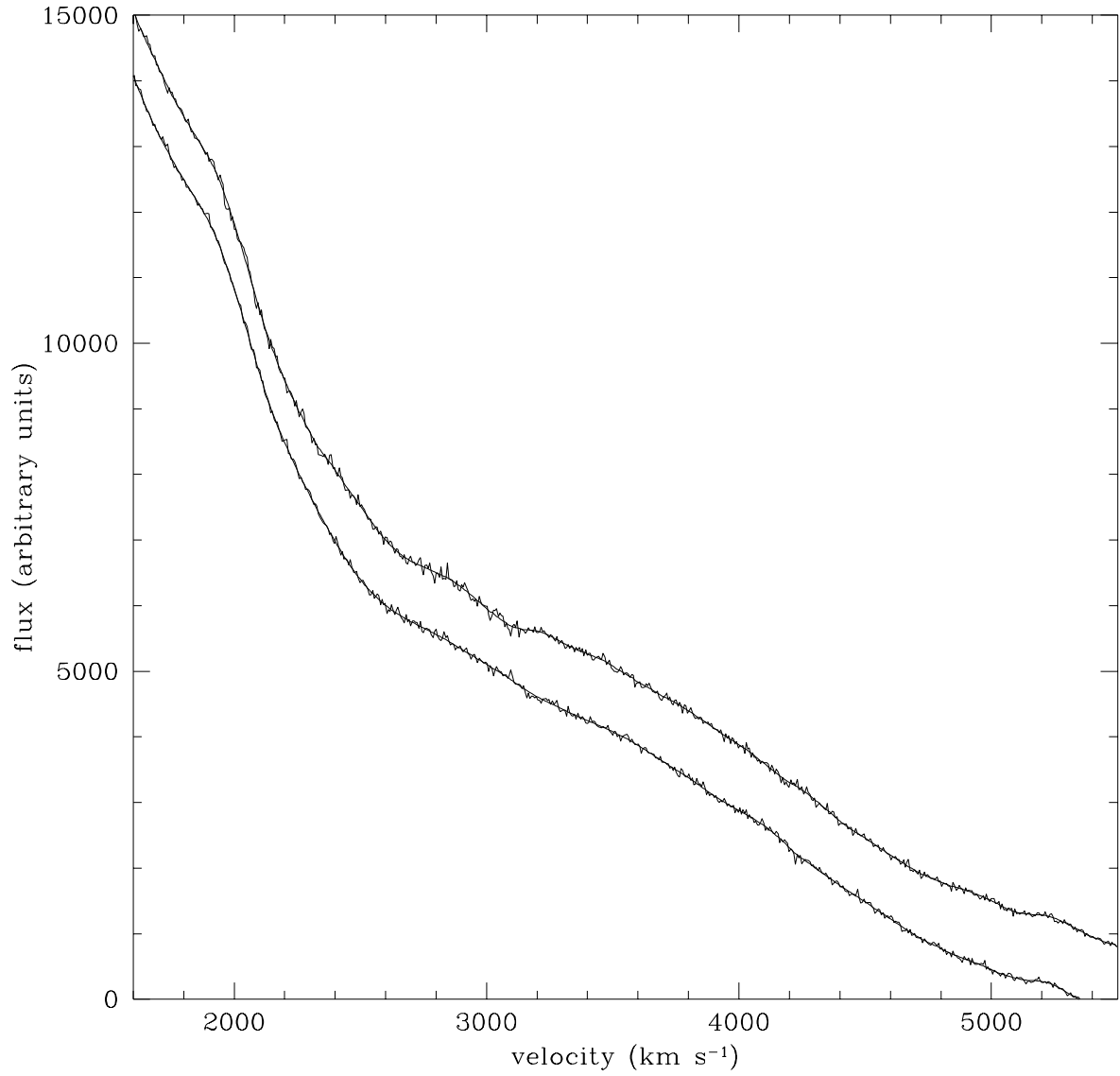}
\caption{
Two different data segments of H${\alpha}$ which were used to calculate
the CC function shown in figure 3. The smooth lines that thread the
data are the polynomial fits for these segments. }  
\end{figure}

\begin{figure}
\plotone{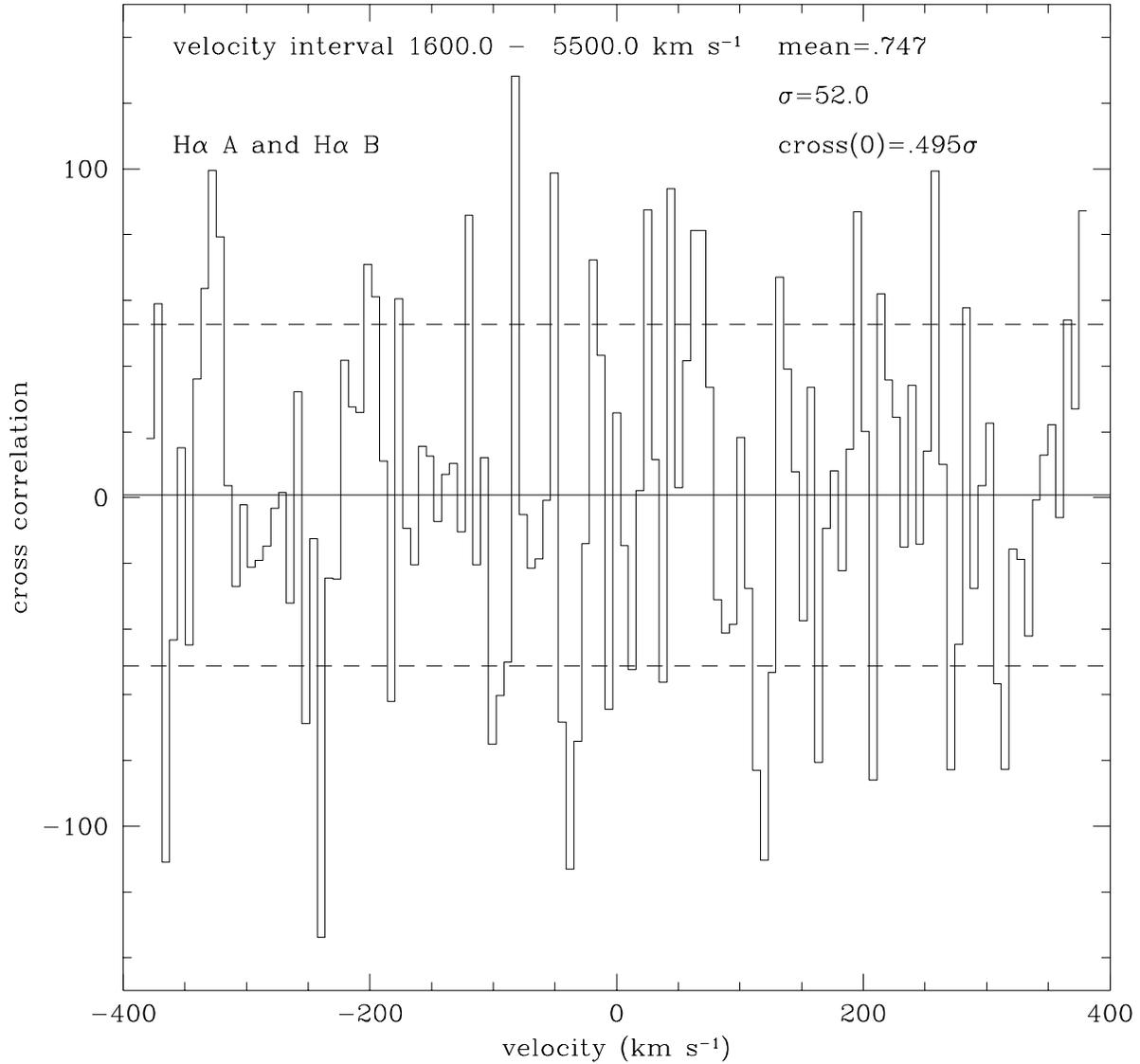}
\caption{Cross correlation (CC) function of the residuals (data--fit) 
 for the data segments shown in figure 2. The CC units are arbitrary.
No significant CC value is evident at zero
velocity shift, which would have been the signature of emission
from discrete units. 
The histogram is the CC function, the
dashed lines are $\pm 1\sigma$, cross(0) is  the value of the function
at zero velocity shift in $\sigma$ units and the solid line is the
function's mean. 
}  
\end{figure}
\begin{figure}
\plotone{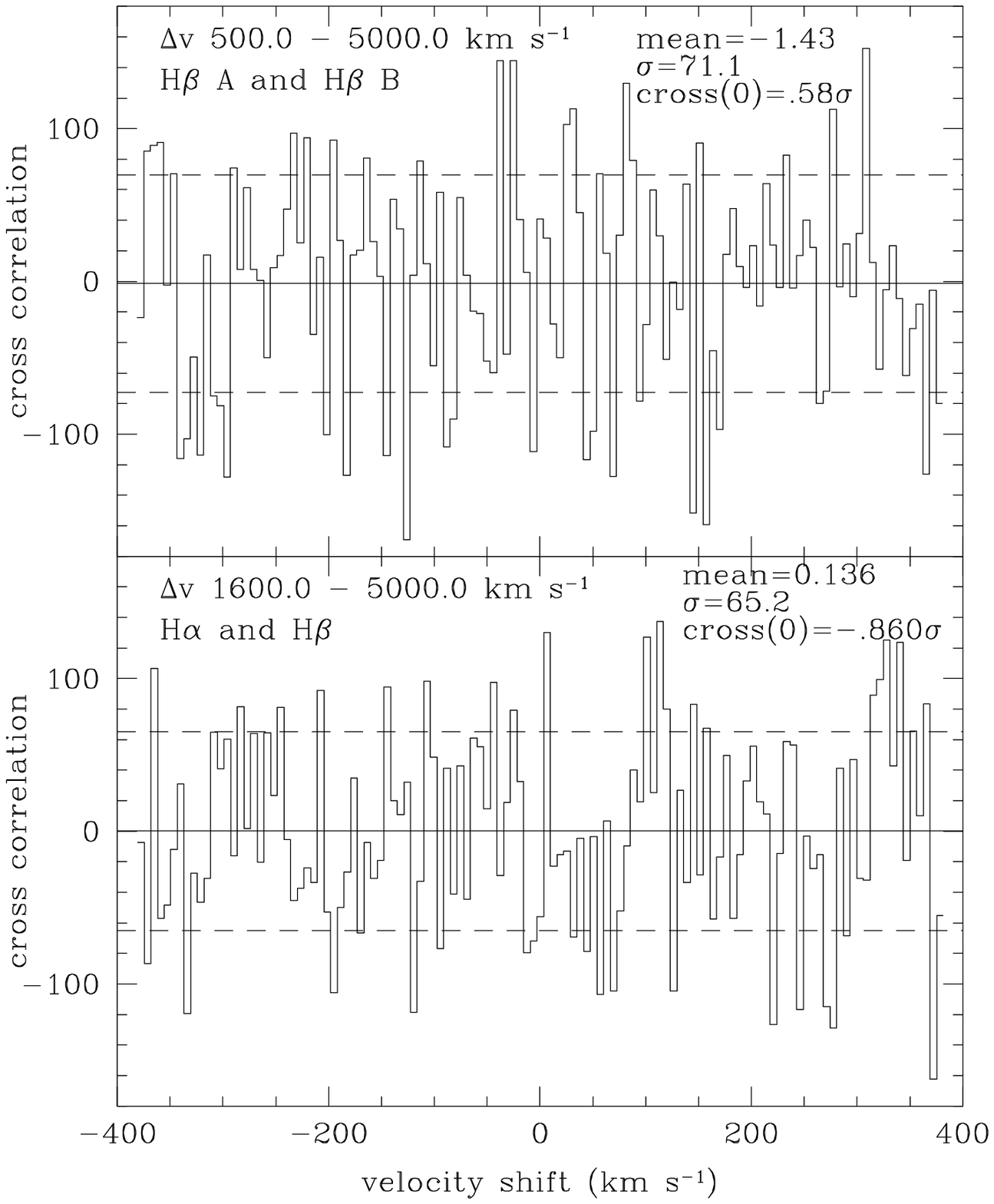}
\caption{Cross correlation function for H${\beta}$ A and H${\beta}$ B 
(upper panel), and for the total data of H${\alpha}$ and H${\beta}$
(lower panel), similar to figure 3.
Again no significant CC  is seen at zero
velocity shift.}  
\end{figure}

\section{CROSS CORRELATION ANALYSIS}

For the CC analysis we used the same algorithm used in Arav (1997).
First we fit the same velocity segment of two line profiles 
with 8th order polynomials. Figure 2 shows the two H${\alpha}$ data 
segments and fits that are used for calculating the CC function shown
in figure 3. Each data segment is composed of two different exposures
and the segments are labeled A (Fig. 2 top) and B (Fig. 2 bottom) 
in figures 4 and 5.
The reduced $\chi^2$ for the fits  were 1.01 (A) 
and 0.98 (B). This assures that systematic
residuals are small (otherwise $\chi^2$ would be significantly larger than 1), 
and that the fit is not too good ($\chi^2$ significantly less than 1).
Since the given errors are purely due to photon
counting statistics, the reduced $\chi^2\simeq1$ result
guarantees that on small scale the errors are governed by
photon shot noise. A close look at the data in figure 2, in which the scale is
magnified by a factor of three compared to figure 1, shows the
similarities and differences between two separate observations
of H${\alpha}$ (each is made of two exposures). The profiles were
first normalized to approximately the same flux level and than shifted
by a small constant flux for ease of comparison.  
Over most of the velocity interval the ratio of the profiles deviates
by less that 1\%. However, in the interval between $2,600-3200$ \km\ the
ratio changes by up to 5\% due to the problems discussed in \S~2.
These fluctuations occur on scales $\gtorder 200$ \km, and thus do not
significantly affect the search for a CC detection on scales of a
few tens of \km. 

Following the definition in Arav (1997), we calculated the CC function of
the residuals extracted from figure 2. The result is shown in figure 3.
We used only the red wing of H${\alpha}$ since the blue wing contains
numerous atmospheric absorption lines which give rise to a strong
CC($\Delta v=0$) value, which is of course unrelated to the signature we
hope to find. We also didn't use the velocity interval $0-1600$ \km,
since it is dominated by the narrow H${\alpha}$ and N~II emission
lines. Two more CC functions are shown in figure 4:
CC function for  the red wing of H${\beta}$ (the blue wing gave a
similar result),
and CC function of the composite observation of   
H${\alpha}$ and H${\beta}$ in a specific velocity interval.
As in the case of Mrk 335, there is no significant CC at zero velocity
shift. In order to try to detect a possible CC on larger scales 
appropriate to realistic clouds (Arav et al. 1997),
 we tried gaussian-smoothing
the CC function with velocity widths of up to 
50 \km. No significant CC was detected on these scale either (as expected
by eye inspection of Figs. 3 and 4).

\section{MONTE CARLO SIMULATIONS}

The lack of a CC signature in the data has
 implications for our understanding of the character and  
disposition of  the line-emitting gas.  We consider this problem in some 
generality by assuming, in turn, that the gas is organized in fundamental 
units of increasing dimensionality and that these elements are distributed in 
velocity space in an essentially uncorrelated manner.

\subsection{Clouds}

This is effectively the zero dimensional case and is a feature of most
published models (e.g.,  Krolik, McKee \& Tarter 1981;  Netzer 1990) 
We suppose 
that the gas is localized, in the form of individual clouds characterized by a 
velocity width $\delta v$ at least as large as the ion thermal speeds at a 
temperature $\sim10^4$~K ($\sim10$~km s$^{-1}$ for hydrogen).    In most 
models, the clouds are presumed to be organized as part of a general, large 
scale flow, with small filling factor; however on velocity scales intermediate
between $\delta v$ and the full velocity width of the line, it is 
reasonable to suppose that the one dimensional velocity distributions, 
comprising clouds drawn from the whole emitting volume will have 
essentially uncorrelated velocities, even if, locally, they do have some 
correlations.  

We used Monte Carlo simulations to check how many clouds are needed in
order not to create a statistically significant  CC($\Delta v=0$)
value. The minimal $N_c$ needed for that is our estimate for the
lower limit on the number of clouds NGC 4151.
The simulations are very similar to those we used in the case of Mrk
335 (Arav et al. 1997), and better understanding of systematic effects
allows us to obtain lower limits up to twice as large as was
previously possible for a given simulation. 
In order to obtain the best lower limit on $N_c$ we simulated the 
H${\alpha}$ profiles, since the H${\alpha}$ data has the highest S/N. 
To have the simulation mimic the real CC analysis, 
we simulated the line segments shown in figure 2. These line
segments cover only about one third of the total flux in the H${\alpha}$
BEL. Once the lower limit for $N_c$ in this segment is established, 
we multiply it by a factor of three to
obtain the
global $N_c$ needed to cover the whole line. It is this number that we
present on the CC plots and use in our discussions.

\begin{figure}
\plotone{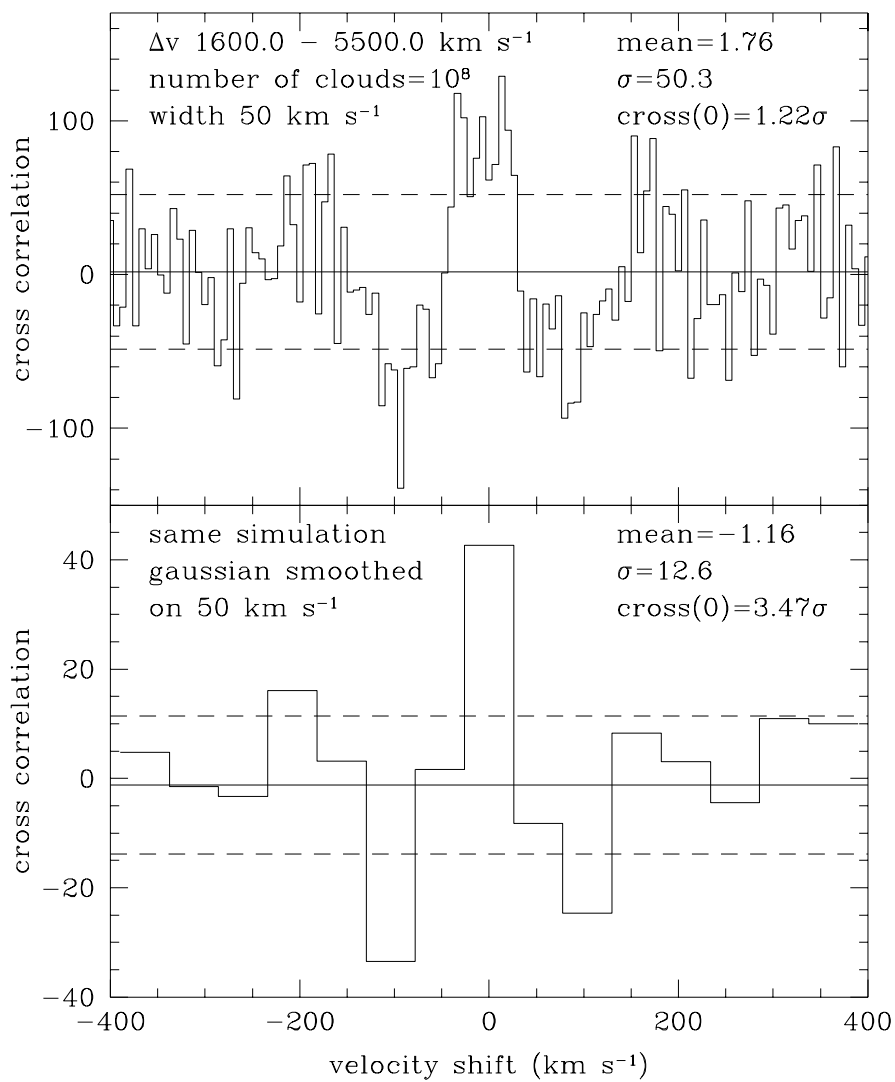}
\caption[]{Top panel: CC function for simulated
H$\alpha$ profiles in a  
 presentation similar to figure 3. The line profile is made from a
distribution of  $10^8$ clouds  where $dN_c/dL\propto L^{-1}$ (see text).
Bottom panel: the same simulation where
the CC was smoothed on a scale corresponding to the width of the
clouds. The significant CC($\Delta v=0$)  from this cloud
distribution is evident.}
\end{figure}

In addition to simulations based on identical clouds we tried cloud
distributions. 
We tested distributions in which $dN_c/dL\propto L^\alpha$ where $L$ is
the luminosity of the cloud. An example in which $\alpha=-1$ (equal total
 clouds' flux in each luminosity decade) is shown in figure 5. The
luminosity interval spreads over three decades and the cloud width was
chosen to mimic realistic clouds, $b=50$ \km. The top panel 
shows the CC function in a similar presentation as in figures 3 and 4.
 Although CC($\Delta v=0$) is not much
larger than 1 $\sigma$, it is evident that a strong CC exists on a
 velocity scale larger than the size of a single resolution element
(6.3 \km), since the CC values for the 10 resolution elements around $\Delta
v=0$ are all larger than 1 $\sigma$. As expected, when we smooth the
CC function on  the velocity scale of the clouds' width, a very
significant CC($\Delta v=0$) emerges (bottom panel). The main reason
that the CC($\Delta v=0$)
 in  this luminosity distribution can be detected even when  
$N_c\gtorder10^8$ clouds, is that the CC is dominated by the more
luminous clouds, since CC $\propto L^2$ and $dN_c/dL\propto
L^{-1}$ in this simulation. 

\begin{figure}
\plotone{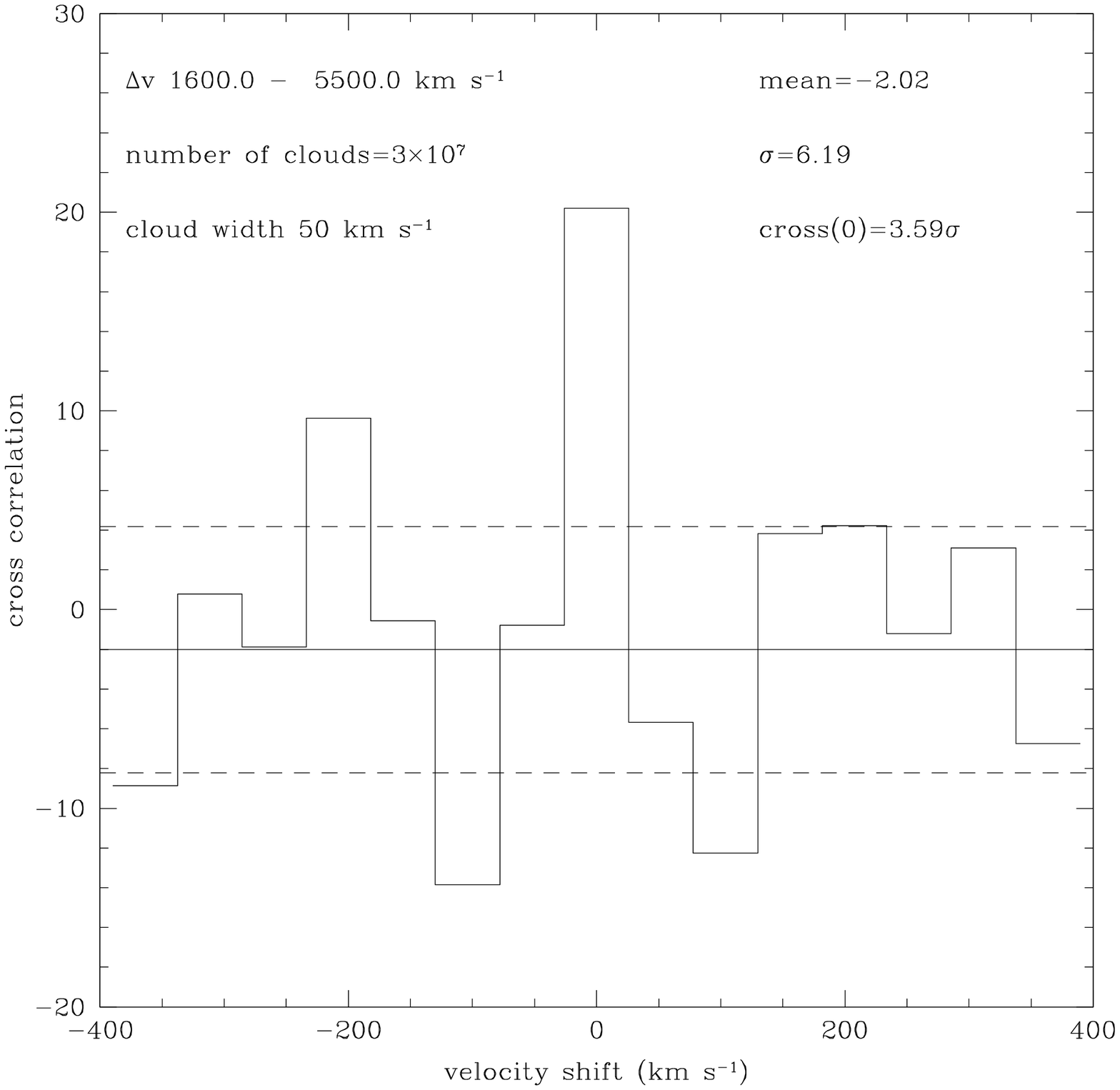}
\caption[]{Smoothed CC function for a simulated H$\alpha$ profile made
of $3\times 10^7$ identical clouds, similar to figure 5.
}  
\end{figure}

In figure 6 we show the CC function for the conservative case of
identical clouds.  This is the case
we use as the firm lower limit for $N_c$ in NGC 4151.

\subsection{Flow Tubes}

Magnetic models (e.g., Rees 1987; Emmering, Blandford \& Shlosmann
1992, Bottorff et al. 1997)
)
posit that the emitting gas is localized on individual one dimensional flux 
tubes, again with a small volume filling factor.  
The gas is confined laterally 
by magnetic stress, but is free to flow along the field lines, 
satisfying a one 
dimensional equation of continuity. The velocity width of the line profile 
from an individual stream is a significant fraction of $\Delta v$, 
but the full 
line is formed  by the superposition of many such profiles. For geometrically 
simple flows, like smoothly accelerating radial motion, the full line profiles 
will naturally be smooth as we observe.  However, more realistic velocity 
fields, for example as is appropriate when the gas originates from a rotating 
accretion disk, give sharp edges to line profiles from individual
streams, which may be observable.

Let us illustrate this possibility by supposing that the gas flows along a 
segment of a circle with uniform speed, 
and constant emissivity per unit length.  The individual 
profile will have the form
\begin{equation}
F_v(v)\propto (\vi^2-v^2)^{-1/2};\qquad v\le \vi .
\label{eq:tube}
\end{equation}
Higher order singularities are possible, but this is the most important and 
prevalent case.  Not all streams contain an extremum in the radial velocity, 
but,  for example, in the solutions computed by Emmering et al (1992),
we estimate 
that about a half of the emission is from gas that will have enough circular 
motion to make this happen. In actuality this line profile must be convolved 
with a local velocity smoothing function that takes account of the thermal 
width and velocity gradients across a flux tube. 
Assuming the individual flux tubes have a characteristic widths to
those derived for the cloud model and that their length is comparable
to the the size of the BLR, then their number must be smaller than $10^4$. 
In order to interpret our 
results in this context, we again constructed Monte Carlo
simulations. 
 We used
flux tubes in the form of quarter circles, where the velocity is
constant along the tube (i.e., $v=v_{\phi}=$constant), and where the
total flux of a given tube is constant. Line profiles
were produced in a similar manner to the cloud simulations, with the
addition that the positional angles of the flow tubes, with respect to
the observer were chosen
randomly.  
We also convolved the emission of a single flux tube with a thermal Gaussian 
($b=20$ 
\km). An interesting result of the flux tube simulations is the
ease of detection of such structures, even for large velocities.
For $v_{\phi}=500$ \km\  $3\times10^7$ flux tubes gave a clear CC
signature, and $10^7$ were clearly detected for $v_{\phi}=1000$ \km.
The ease of detection stems from the projected emission profile seen
by the observer. Flow tubes can give rise to cusps in the emission
profiles, which are very narrow compared to the nominal velocity of
the flow. As discussed above, the CC is proportional to the square of
the flux at a given velocity position. Thus, relatively few cuspy
profiles can dominate the  CC function. In addition, a face on
view of the plane of the tube give rise to emission concentrated in a
small $\Delta v$, which leads to a similar effect.

\subsection{Shock Fronts and Expanding Shells}

The next kinematic possibility to consider is a two dimensional
surface, epitomized by 
a radiative shock forming in the outflow perhaps enveloping an
obstacle or in response to variation in the initial flow velocity 
(Perry \& Dyson 1985). 
Again, we suppose that the emitting gas occupies a 
small fraction of the volume but that there are a large number  of these
curved surfaces.

Our elementary model will be a single spherically expanding surface with 
constant radial velocity $\vi$ and constant surface 
emissivity.  The observed line profile is given by
\begin{equation}
F_v(v)=\mbox{const}. \qquad v\le \vi.
\label{eq:shell}
\end{equation}
 We added a pure thermal broadening and found that
for shells expanding at 100 \km\ with a thermal broadening of $b=25$ \km,
a CC($\Delta v=0$) is clearly detected for $3\times10^6$ shells.  For
similar shells expanding at 200 \km\ the limit is $10^6$. The
detectability also depends on the gradient of flux as a function
of velocity.  The smaller the gradient (wider smoothing) 
at the edges of the profile, the harder it is to detect the CC signature.  

In figure 8 we
show the different emission profiles we used in the simulations. The
three profiles at the center of the figure are due to randomly
oriented flux tubes with a flow velocity of 1000 \km, and the
same total flux. An edge on view of the tubes' plane correspond to
$\theta=0$, and $\phi$ is the starting position of the tube, where
$\phi=0$ is along the line of sight. A near face on view is seen in
the  $\phi=323$, $\theta=100$ case, and a cuspy emission profiles is shown 
in the $\phi=62$, $\theta=1.5$ case.  A real discontinuity is seen in
the cuspy case, which occurs because emission from two different
locations along the tube
contribute at the same velocity. Since the tube is finite, at some
point one contribution ends, leading to the discontinuity.
In figure 8 we did not smooth the emission profile of the flux tube in
order to visually enhance the kinematical effects. In the simulations
we smoothed the emission profiles with $b=20$ \km.
An expanding shell emission profile and the clouds' gaussian profiles 
  (centers shifted to $-750$ \km\ and $900$ \km, respectively,  for
presentation purposes) are also shown in figure 8.

\subsection{Space Filling Flows}

Finally, for completeness, we consider flows that are stationary and 
continuous, filling space over a limited volume in the AGN.  A good example 
is the model of Chiang \& Murray (1996).  The flow is, by definition, smooth 
with a velocity field which must reproduce the observed line profile.  No 
constraint on this model is imposed by our observations.

\begin{figure}
\plotone{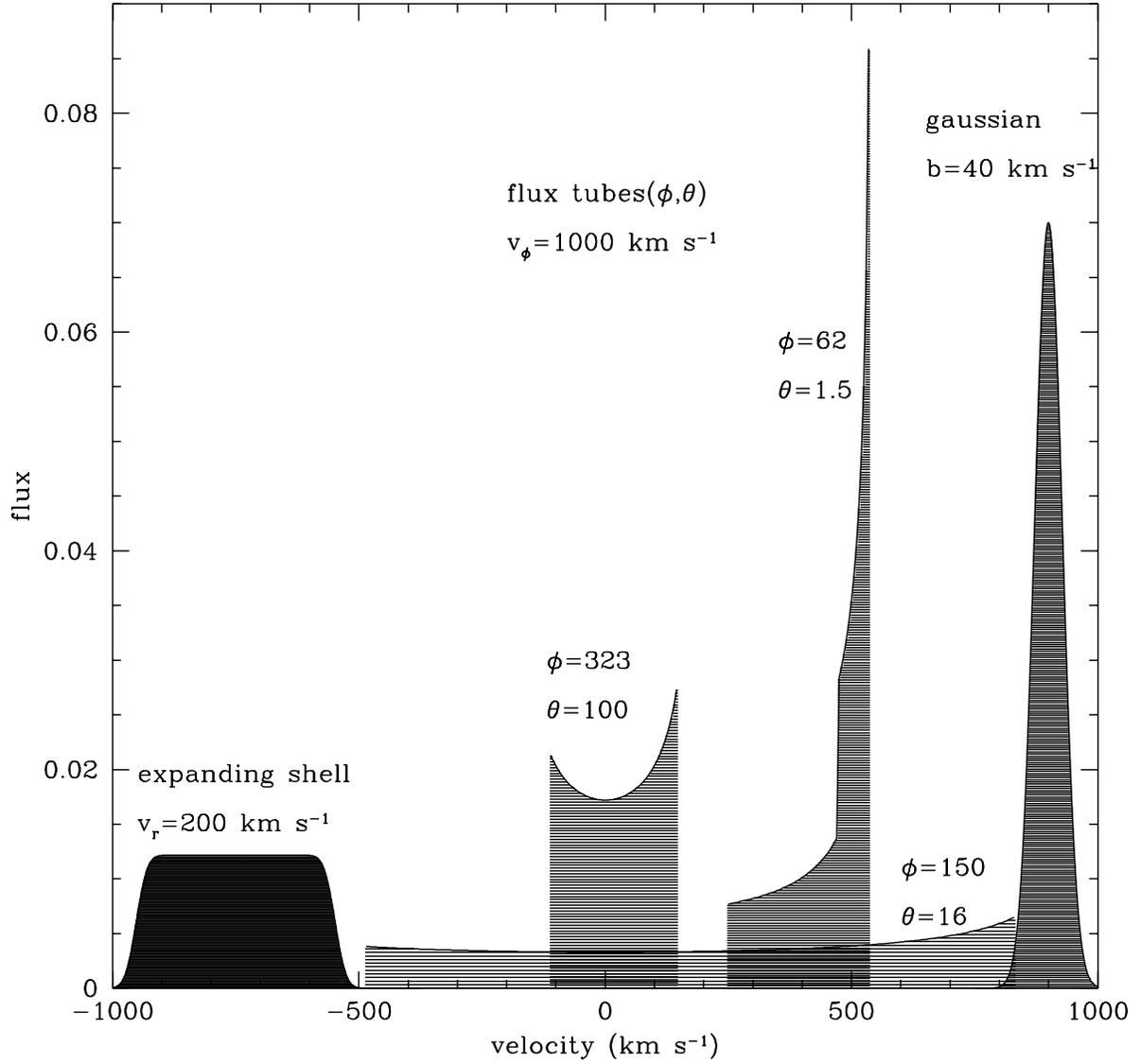}
\caption[]{Emission profiles used in the Monte Carlo simulations (see \S~4).
}  
\end{figure}

\begin{figure}
\plotone{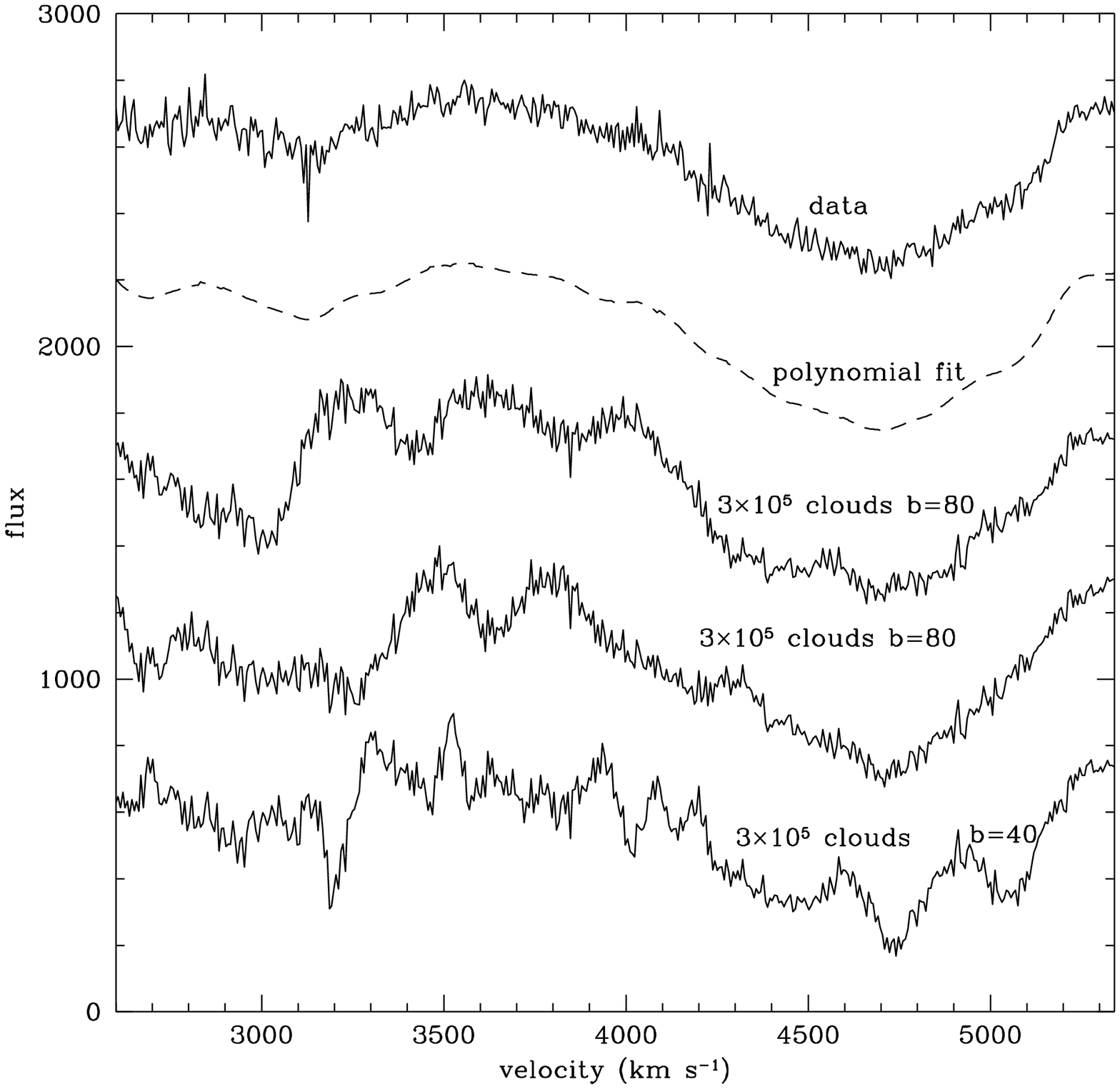}
\caption[]{A qualitative comparison between the observed H$\alpha$
line (top curve) and simulated ones (bottom three curves) 
shows how difficult it is to create a
smooth enough profile from $3\times10^5$ clouds even for very broad
individual cloud emission. All the curves are shown on the same scale
which is five times magnified in the y axis and 1.42 in the x direction
with respect to the H$\alpha$ segment shown in figure 2.
 A linear function was
subtracted from the profiles in order to obtain a flat presentation.
The second curve shows the polynomial fit for the data that was used
in creating the simulated profiles. 
 }  
\end{figure}

\section{DISCUSSION}

Our observational result appears to be in conflict
with the simple BEL cloud model, and by extension also with flux tubes
and expanding shell models. 
One possible explanation for the lack of persistent microstructure in the line
profiles is that this structure exist but changes on timescales
comparable with or shorter than the time it took to acquire the
observations. This possibility is unlikely, since in that amount of
time the velocity change 
 for the clouds is too small. 
 The acceleration of the clouds in NGC 4151 can be estimated as
follows: 
From reverberation mapping, 
Maoz et al. (1991) estimated
the size of the BLR to be nine light days or $R\sim 2\times10^{16}$ cm. 
The velocity of the clouds is given by the width of the BELs $(\sim
5\times 10^8$ cm s$^{-1}$. Thus, $a\sim  v^2/R\simeq 10$ 
cm s$^{-2}$. In the few hours it took to obtain the observations the
clouds should therefore accelerate by roughly 1 \km. Since the CC
analysis should be  able
to detect structure on a scale of $\sim50$ \km (see Figs. 5 and 6) the
change in velocity of the clouds due to acceleration should be
negligible.
    
In addition we note that our underlying assumption in the simulations 
is that the clouds' velocity distribution is Poissonian.  
It is reasonable to expect that
if the clouds' motion is highly ordered on small velocity scale, the resulting
CC signature will be easier to detect.  It is difficult to imagine
a noncontrived way in which an ordered flow gives a smaller CC
signature than the  Poisson distribution we use.  However, if such 
an ordered distribution exists the limits set by this analysis would be
weakened accordingly.

We already showed that the current bloated star models are ruled out by
our lower limit on $N_c$ in Mrk~335 (Arav et al. 1997). 
The NGC~4151 results strengthen 
this argument considerably since our lower limit for $N_c$ has risen
by an order of magnitude and since the expected number of bloated
stars is smaller in this less luminous object. Unless the emission
width from an individual star is significantly larger than 100 \km,
the lack of stable micro structure in the line profiles strongly
rules out all bloated star models.  
Even
without resorting to the CC analysis it is easy to demonstrate that 
for reasonable cloud's emission
width,
the BELs in NGC 4151 cannot be made from $10^4-10^5$ individual
emitters, which is the maximum allowed number of bloated stars. 
 Figure 7 shows part of the H$\alpha$ data from the combined
exposures, and three simulations of the same segment. 
We have used the same noise but different cloud distributions in
each simulation.
The amount of structure in the simulations, as contrasted with
the smoothness of the data, shows that either
 the observed H$\alpha$ profile is  made from
many more than $3\times10^5$ clouds, or that the emission width of an
individual cloud is much larger than $b=80$~\km. The structure that is
seen in the data is on scales of $500-1000$~\km\ and is difficult to
associate with individual units of sonically connected material, but
can be associated with a smooth flow or disk models. 
To the eye it also seems that the data has structure on
 scales of $\sim50$ \km\, but the CC analysis have shown that this
structure does not repeat in two different observations.  

From the Monte Carlo simulations, it is clear that strong
constraints can also be applied to emission models that does not arise
from a point like structures (i.e., clouds).
For example,
although we only used one simple choice of a flow tube, some general
conclusions can be deduced from the simulations of this case.  It appears that
the number of randomly oriented flow tubes must be very large in order
to conceal a significant CC($\Delta v=0$).  This result holds even
for flux tubes that on the average span a significant fraction of the
velocity width of the line. In order for a flow tube emission model
not to produce a CC signature one or more of the followings must hold:
a) The number of flow tubes is very large. b) emission cusps in
projected velocity are somehow avoided.  c) The flow tubes are
organized in a way which mimics a smooth ordered flow.  d) The
velocity width perpendicular to the axis of the tube is substantial
($\gtorder 100$ \km) and thus dampens the effect of the narrow emission cusps.

As noted in \S~1, 
photoionization arguments give the highest model-independent (i.e.,
irrespective of the clouds' microphysics)
estimate for $N_c$.  
in NGC~4151 these arguments lead to an upper limit estimate of
$10^6-10^7$ clouds  (extrapolated from 
Arav et al 1997; but see caveats therein and in \S~1).
This upper limit is smaller by about an order of magnitude from the
one in Mrk~335 due to the lower luminosity of NGC~4151, or more
precisely, due its smaller  BLR size. In contrast, our lower limit for
$N_c$
in this object has increased by an order of magnitude to $3\times
10^7-10^8$. While the Mrk 335 result does not put a strong constraint
on the generic clouds picture, our current one does.
The photoionization estimates can be pushed up, but for the price of
becoming more contrived. The constraint our results put on any clouds'
picture is strong enough to suggest that the BELs do not originate
from discrete quasi-static structures, and thus strengthen
the case for emission models that create the BELs from a 
 continuous flow or in the accretion disk.


We thank S. Vogt for
leading the construction of the HIRES spectrograph and Michael Rauch
for assisting with the observations.
N. A. and R. D. B. acknowledge support from NSF grants 92-23370 and 95-29170.
A. L. acknowledges support from the E. and J. Bishop research 
fund, and from the Milton and Lillian Edwards academic lectureship fund.


\begin{references}
     

\reference Alexander, T., \& Netzer, H. 1994, MNRAS, 270, 803 
\reference Atwood, B., \& Baldwin, J. A., \& Carswell, R. F. 1982, ApJ, 257, 559 
\reference Arav, N., Barlow, T., Laor, A., \& Blandford, R. D. 
1997, MNRAS, 288, 1015
\reference Barlow, T., \& Sargent, W. L. W.,  1997, AJ, 113, 136 
\reference Bottorff, M., Korista, K. T., Shlosman, I., \& Blandford, R.
D. 1997, ApJ, 479, 200

\reference  Chiang, J., \& Murray, N. 1996, ApJ, 466, 704
\reference Crenshaw, D. M., et al., 1996, ApJ, 470, 322
\reference Edelson, R. A., et al., 1996, ApJ, 470, 364
\reference Emmering, R. T., Blandford, R. D., \& Shlosman, I. 
1992, ApJ, 385, 460
\reference Davidson, K., \& Netzer, H. 1978, Rev. Mod. Phys, 51, 737 
\reference Ferland, G. J., \& Rees, M. J. 1988, ApJ, 332, 141
\reference Kaspi, S., et al., 1996, ApJ, 470, 336
\reference Kazanas, D. 1989, ApJ, 347, 74 
\reference Krolik, J. H., McKee, C. F., \& Tarter, C. B. 1981, ApJ, 249, 422
\reference Maoz, D., et al., 1991, ApJ, 367, 493
\reference Netzer, H. 1990, in Saas-Fee Advanced Course 20: Active
Galactic Nuclei, ed. R. D. Blandford, H. Netzer,  \& L. Woltjer  
(New York: Springer), 57
\reference Perry, J. J., \& Dyson J. E. 1985 MNRAS, 213, 665 
\reference Rees, M. J. 1987, MNRAS, 228, 47p  
\reference Scoville, N., \& Norman, C. 1989, ApJ, 332, 162 
\end{references}
\end{document}